%% file: TXT_17-TIE-3215.tex
\documentclass[10pt,journal]{IEEEtran}

\usepackage[utf8]{inputenc}
\usepackage{amsmath,amsfonts,amsbsy,amssymb}
\usepackage{mathabx}
\usepackage{mathrsfs}
\usepackage{tabularx}
\usepackage{graphicx}
\usepackage{cite}
\usepackage{wasysym}
\usepackage{multirow}
\usepackage{float}
\usepackage{color}
\usepackage[colorlinks=false,urlcolor=black,hidelinks]{hyperref}
\usepackage[english]{babel}

 \usepackage{tikz}
 \usetikzlibrary{shapes,arrows}
 \usepackage[absolute,overlay]{textpos}
 \usepackage{tikz}
 \usetikzlibrary{topaths}
 \usetikzlibrary{shapes,trees,snakes}
 \usetikzlibrary{arrows}
 \usetikzlibrary[shadows]
 \usetikzlibrary{positioning}
 \usetikzlibrary{matrix}
 \usetikzlibrary{shapes.geometric}
 \usetikzlibrary{decorations.pathmorphing}
 \usepgflibrary{patterns}
 \usetikzlibrary{calc}
 \usetikzlibrary{fit}			
 \usetikzlibrary{backgrounds}
 \usetikzlibrary{decorations.markings}

\usepackage{algorithm}
\usepackage[noend]{algpseudocode}
\usepackage{algorithmicx}

\begin{document}

\title{Long-range Low-power Wireless Networks and Sampling Strategies in Electricity Metering}

\author{
	\vskip 1em
	{Mauricio C. Tomé, \emph{IEEE Student Member},
	Pedro H. J. Nardelli,
and Hirley Alves, \emph{IEEE Member}
	}

	\thanks{
		
		{\textbf{Manuscript accepted and published in IEEE Transaction on Industrial Electronics.}
		This work is partially supported by Aka Project SAFE (Grant n.303532) and Strategic Research Council/Aka BCDC Energy (Grant n. 292854).		The authors are with the Centre for Wireless Communications (CWC) at University of Oulu, Finland. 		
		P. H. J. Nardelli is also with Laboratory of Control Engineering and Digital Systems, Lappeenranta University of Technology, Finland. Contact: pedro.nardelli@lut.fi.
		}
	}
}

\maketitle
\vspace{-1.8cm}
%
\begin{abstract}
This paper studies a specific low-power wireless technology capable of reaching a long range, namely LoRa.
Such a technology can be used by different applications in cities involving many transmitting devices while requiring loose communication constrains.
We focus on electricity grids, where LoRa end-devices are smart-meters that send the average power demanded by their respective households during a given period.
The successfully decoded data by the LoRa gateway are used by an aggregator to reconstruct the daily households' profiles.
We show how the interference from concurrent transmissions from both LoRa and non-LoRa devices negatively affect the communication outage probability and the link effective bit-rate.
Besides, we use actual electricity consumption data to compare time-based and event-based sampling strategies, showing the advantages of the latter.
We then employ this analysis to assess the gateway range that achieves an average outage probability that leads to a signal reconstruction with a given requirement.
We also discuss that, although the proposed analysis focuses on electricity metering, it can be easily extended to any other smart city application with similar requirements, like water metering or traffic monitoring.
\end{abstract}

%

\section{Introduction}
The Internet of Things (IoT) denotes the widespread deployment of communication networks between machines without direct human intervention \cite{Manyika2015TheHype}.
One important IoT application is in the monitoring of different aspects within cities to help their complex and distributed management \cite{Zanella2014InternetCities}.
Household management of electricity consumption is a good example case, where IoT allows for a better understanding of how the electricity is consumed, potentially indicating ways to improve its usage efficiency in both individual and aggregate levels \cite{Siano2014DemandSurvey}.

Beyond this, IoT allows for regulatory actions into the physical system based on information \cite{Kuhnlenz2016DynamicsLayers}.
For instance, the utility -- informed by IoT devices -- may directly control few households' heating devices -- via IoT devices -- aiming at monetary savings \cite{Palensky2011DemandLoads}.
Appliances connected to the same network may directly coordinate their reactions in respect to the grid frequency to help balancing supply and demand (e.g. fridges postponing or anticipating their cycles without creating more instabilities \cite{Evora2015SwarmGrids}).

The term IoT, however, is also very broad, covering extreme application cases: from massive deployment with loose requirements (e.g. air quality measurements) to very specific high-reliability low-latency applications (e.g. robot arms in fully autonomous industrial plants) \cite{Raza2017LowOverview}.
In this paper, we target an application with relatively loose requirements related to the communication system, namely electricity metering in households.
Specifically, the scenario under investigation consists in a typical household that sends its average demanded power during a given period to an aggregator node (which can be the utility or a micro-grid trader, depending on the distribution arrangement in question) \cite{Nardelli2016MaximizingConstraints,Tome2016JointUsers}.
At the end of each day, the aggregator wants to reconstruct the power demand curve from the successfully decoded samples transmitted by the household.
The aggregator may use this information to, for example, plan for the next day operations (e.g. \cite{Ma2017TheNetworks}) or profiling consumers (e.g \cite{Li2017LoadDomain}); although the use actually defines the required quality of service provided by the communication network (e.g. \cite{Dawy2017TowardCommunications}), we keep it here unspecified and only look at the relation between the performance of the communication link and the quality of the signal reconstruction.

As to be explained throughout this paper, our contribution to the topic is the following.
\begin{itemize}
	\item We extend the Long Range (LoRa) technology study case from \cite{Georgiou2017LowScale} by (a) including in their proposed stochastic geometry model a density of external interferers, and (b) using a random spreading factor allocation, which is fairer in relation to our specific application.
    \item The theoretical model developed in \cite{Nardelli2016MaximizingConstraints} is modified to incorporate realistic LoRa setups from \cite{Georgiou2017LowScale} (which is supported by actual deployments as discussed in \cite{Raza2017LowOverview}.
    \item We reproduce the sampling strategies from \cite{Tome2016JointUsers} and \cite{Simonov2014HybridGrid} by testing their proposed approach in a different dataset \cite{Zimmermann2009End-useSavings} and providing further comparisons between the time-based and event-based schemes. As in \cite{Tome2016JointUsers}, we show that, for most households evaluated, the latter option consistently provides lower reconstruction errors when the number of samples are similar.
    \item We investigate the trade-off involved between the system variables pointing out what should be the most suitable gateway range to achieve a communication outage probability that can sustain a signal reconstruction by the aggregator within a given quality level.
\end{itemize}

The rest of this paper is divided as follows.
Sec. \ref{sec:related} provides a short overview of the relevant literature about LoRa and stochastic geometry applied in wireless networks, mainly discussing our main advancements.
Sec. \ref{sec:system} contains the system model based on stochastic geometry and metrics to evaluate the performance of the proposed LoRa deployment, as well as the respective numerical results.
We present the sampling strategies and their assessment in Sec. \ref{sec:sampling}.
Sec. \ref{sec:discussions} discuss about the implications of these results focusing on deployment aspects in smart cities, while Sec. \ref{sec:conclusions} concludes this paper, also listing possible research directions.

\vspace{-1ex}
\section{Related work}
\label{sec:related}
\subsection{Long Range Technology -- LoRa}
Low power wireless technologies covering wide areas are becoming trendy nowadays, as indicated by the survey \cite{Raza2017LowOverview}.
The reason for that is the potential to reach the massive number of IoT devices at low cost with reasonably efficient performance.
It is worth saying, however, that the so-called ``low-power wide-range'' technologies are well suited to the massive machine-type Communications (MTC) in contrast to the critical MTC, whose applications require (very) low latency (1-10 ms) and (ultra) high reliability (99.999\%) \cite{Popovski2014Ultra-ReliableSystems}.
The main advantages of Low Power Wide Area (LPWA) technologies are, beyond the long range and the low power consumption themselves, their low cost and scalability while guaranteeing some (not so strict) quality of service.
For some specific applications as presented in \cite{Dawy2017TowardCommunications}, LPWA may become a way to alleviate the data traffic in traditional cellular networks to avoid some problems presented in \cite{Madueno2016AssessmentGrid}.

Among other options discussed in \cite{Raza2017LowOverview}, Long Range (LoRa) -- a proprietary technology proposed by Semtech and promoted by the LoRa Alliance -- provides bidirectional communication based on chirp spread spectrum modulation that spread the narrow-band signals over a wider bandwidth.
LoRa uses a star-of-stars topology consisting of end-devices (in our case, smart-meters), gateways and a central network server.
The end-devices directly send their message via a wireless channel with six different spreading factors (SF7 to SF12) using unslotted Aloha as its medium access control; in Europe the spectrum used by LoRa is in the 863-870 MHz ISM Band range with channels of bandwidth of 125 kHz. 
The communication from the gateway to the server occurs through non-LPWA networks (e.g. cellular or Ethernet).
End-devices are divided into three classes (A, B and C) that are most related to their downlink capabilities.
Aspects relevant to this article, such as data-rate and successful detection at the uplink, will be given in the next section, while more general details about LoRa can be found at \cite{Raza2017LowOverview}, \cite{Adelantado2017UnderstandingLoRaWAN} and references therein.

\vspace{-1ex}
\subsection{Stochastic geometry for wireless networks}
Communication engineering communities have quite established analytical frameworks to account for channel and/or traffic uncertainties, which are the basis of almost all (if not all) wireless communication systems \cite{Yacoub1993FoundationsEngineering}.
The uncertainties related to the relative positions between devices are, however, still under development; positions have been traditionally modeled as regular grid-like topologies (e.g. hexagonal or square cellular networks) or toy-models with few communication devices (e.g. two-hop systems).
As a growing research field, stochastic geometry and spatial point process theories applied in wireless networks (e.g. \cite{Haenggi2013StochasticNetworks}) are able to capture many trade-offs involved in the system design and deployment explicitly including the uncertainties related to the devices' positions.
Their main advantage is that the aggregate interference becomes analytically treatable, so neither time-consuming Monte Carlo simulations nor (over-)simplifying assumptions about the aggregate interference (e.g. completely neglecting it) are needed in order to evaluate the system performance.

Our contribution here is mainly based on three papers that employ such a model.
In \cite{Georgiou2017LowScale}, the authors studied whether LoRa can scale considering Spreading Factor (SF) allocations based on distance.
Their results indicate a exponential dependence of the number of end-devices and the outage probability, which is the complement of the probability that a given message is correctly received by the gateway.
They show that success probability exponentially decays with the average number of devices, evincing the negative effects of interference in LoRa even with its mitigation techniques.
But, the most interesting aspect of this paper is the stochastic geometry treatment of interference.
Assuming that end-devices follow a Poisson point process and making use of order statistics, the authors found a closed-form expression for the probability that an outage event being caused by another concurrent end-device using the same SF.
This outage event is defined when the received power of a given desired signal is not four times (i.e. 6 dB) stronger than any other concurrent transmission in the same SF.
In any case, the effect of the aggregate interference from all users using the same SF is not considered, but only the effect of the dominant (stronger) interferer.

In the other two articles \cite{Nardelli2016MaximizingConstraints,Tome2016JointUsers}, the authors propose a general optimization of the link throughput based on Shannon capacity, where smart meters and aggregators (acting as a base-station, a gateway) are unlicensed users of the uplink channel of the cellular network in a spectrum sharing setting.
Their proposed scenario only considers the effects of the interference from the licensed mobile devices (modeled as a Poisson point process) in the aggregator reception, showing that the maximum achievable throughput actually happens with a less strict outage constraint.
This higher outage constraint, however, is shown to have a small effect in the average power demand signal reconstruction by the aggregator. 
We find, nevertheless, that these results are still very analytical, only relying on an abstract conception of cognitive radio and spectrum sharing without specifying any possible technology.

\vspace{-2ex}
\subsection{Sampling strategy}
Another interesting result from \cite{Tome2016JointUsers} is the comparison between time- and event-based sampling, indicating the first may lead to redundant samples for most of the day, while not good enough during other times.
The idea of event-based sampling for sensor networks has been discussed for some time (e.g. \cite{Miskowicz2006Send-On-DeltaStrategy}, but only very recently adopted in electricity metering scenarios (e.g. \cite{Simonov2014HybridGrid,Simonov2017GatheringMetering}).
However, only in \cite{Tome2016JointUsers} the joint performance between sampling and the link outage probability is analyzed (but with the previously discussed limitations).

\subsection{Relation to this contribution}
This paper is mainly built upon \cite{Georgiou2017LowScale,Nardelli2016MaximizingConstraints,Tome2016JointUsers}.
In relation to \cite{Georgiou2017LowScale}, we extend the stochastic geometry modeling by considering outage events due to not only the dominant interferer, but also the aggregate interference in a given SF.
Besides, instead of outage caused by Gaussian noise, we consider the aggregate interference of non-LoRa devices by explicitly including a density of interferers that transmit in the same channel.
We then have another variable affecting the outage probability when the non-LoRa interference is treated as noise \cite{Nardelli2015ThroughputInformation}.

This approach follows \cite{Nardelli2016MaximizingConstraints}, but specifies it to an actual LoRa deployment.
In other words, we move from Shannon capacity in bit/s/Hz and their abstract conceptual model (even including highly directional antennas that leads to negligible unlicensed-to-licensed and unlicensed-to-unlicensed user interference).
We analyze here a LoRa scenario following the system setting from \cite{Georgiou2017LowScale}, which is based on the technology specification and several deployment trials (refer to \cite{Raza2017LowOverview} and references therein).
Besides, the time-based and event-based schemes employed in \cite{Tome2016JointUsers} are implemented and tested for another electricity consumption database \cite{Zimmermann2009End-useSavings}, which has different granularity and household composition as well as geographical location.

\vspace{-1ex}
\section{LoRa deployment}
\label{sec:system}
\subsection{System model and performance metrics}
\begin{figure}[!t]
\resizebox{1.0\columnwidth}{!}{\input{lorafig.tikz}}
\caption{{Illustrative figure of the network deployment where LoRa end-devices (smart meters in this paper) are the black and blue circles, while the LoRa gateway is the square node at the center. The black circle linked by a dashed line to the gateway is the reference link, located $r$ km away. The other black circles represents LoRa devices using the same spreading factor (SF) as the reference link, and the blue ones are LoRa devices  using other SFs. LoRa devices cause interference to each other only if they are transmitting at the same time with the same SF (represented here by the black circles). Note that LoRa devices may also suffer interference from non-LoRa devices (depicted by red circles). Since these devices use different radio access technology, but in the same frequencies, they are treated as noise by the LoRa gateway.}}
\label{fig:deployment}
\end{figure}
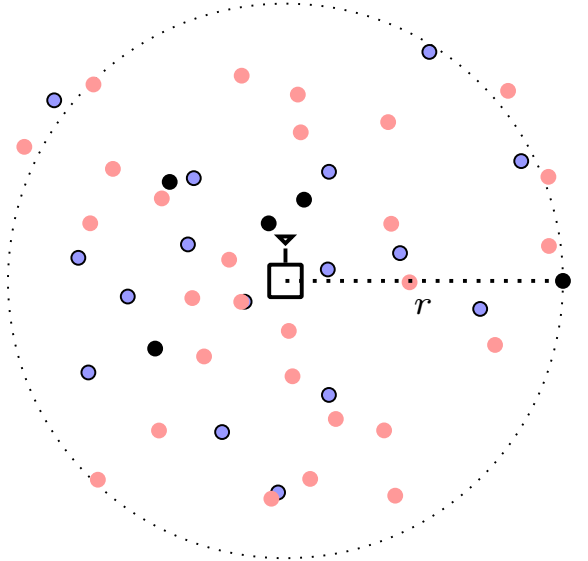

Fig.\ref{fig:deployment} depicts an illustrative network deployment. 
We analyze here the LoRa deployment from \cite{Georgiou2017LowScale} with some differences to be explained in the following.
We assume that smart-meters need to transmit to a single gateway a 25 byte message in a bandwidth of 125 kHz containing the cumulative energy consumption in a given period of time as well as the period.
The meters' locations are randomly distributed in the plane following a Poisson point process $\Phi_\mathrm{SF}$ with density $\lambda_\mathrm{SF}$ devices/km$^2$ \cite{Haenggi2013StochasticNetworks}.
When a given smart-meter wants to send a message to the gateway, the LoRa network server assigns to it a spreading factor (SF), which is randomly and independently allocated from SF7 to SF12.
So a smart-meter has a probability of $p_\mathrm{SF}=1/6$ to be allocated to a specific SF, which is independent across the points from $\lambda_\mathrm{SF}$.

The smart-meter traffic is  related to the sampling strategy, which will be discussed later in Section \ref{sec:sampling}.
Regardless of the strategy adopted, the actual transmission is randomized to decrease the number of concurrent transmissions.
A duty cycle limitation of 1\% must be also assumed so that each smart meter has a limited number of wireless transmissions per day.
The actual density of smart-meter concurrently transmitting is  then (much) smaller than $\lambda_\mathrm{SF}$.
We assume the density of active smart-meters as $p_0 \lambda_\mathrm{SF}$, where $p_0$ is the probability that any smart-meter from $\Phi_\mathrm{SF}$ is active while a given reference link is also transmitting.
Using the point process theory nomenclature, the mapping from $\lambda_\mathrm{SF}$ to a process in which some of the original points were erased is known as \textit{thinning}.
As the transmissions use LoRa, an outage at the reference link occurs when the other active smart-meters using the same SF lead to a signal-to-interference ratio (SIR) at the gateway in respect to the reference link below a given threshold $\beta_\mathrm{SF}$.
As we assume a random SF allocation with probability $p_\mathrm{SF}$, another thinning happens, resulting in a density $\lambda = p_\mathrm{SF} p_0 \lambda_\mathrm{SF} $ of active smart-meters using the same SF.

We consider that the wireless channel is composed of two components: a distance dependent path-loss with exponent $\alpha>2$ and a gain due to multipath. 
We assume the basic path-loss equation, where the received power is proportional to $r^{-\alpha}$ with $r$ being the distance between the smart-meter and the gateway, while the multipath is modeled as independent and identically distributed channel gains related to a Rayleigh fading distribution with unity mean.
Then, we can compute the outage probability $P_\mathrm{out:1}$ from smart-meter as \cite{Haenggi2013StochasticNetworks,Nardelli2016MaximizingConstraints}:
\begin{equation}
P_\mathrm{out:1} = 1 - \exp\left(-k \lambda r^2 (\beta_\mathrm{SF})^{2/\alpha}\right),
\label{eq:out1}
\end{equation}
where $k = \pi \Gamma(1 + 1/\alpha) \Gamma(1 - 1/\alpha)$ with $\Gamma( \cdot)$ being the Gamma function.

\begin{table}[!t]
	%
	\renewcommand{\arraystretch}{1.3}
	\centering
	\caption{LoRa setting adapted from \cite{Georgiou2017LowScale}}
	\label{tab:LoRa}	
	\centering
	\begin{tabular}{l l l}
		\hline \hline
		\textbf{Spreading factor} $\mathrm{x}$			& \textbf{Bit-rate} $R_\mathrm{SFx}$ in kb/s& \textbf{Minimum SIR}	$\beta_\mathrm{SFx}$\\ 		\hline
		7 					& 	 5.47 	& 	0.25	\\ 
		8 					& 	 3.13   & 	0.125	\\ 
		9					& 	  1.76   & 	0.06	\\
        10					& 	 0.98    & 	0.03	\\
        11					& 	  0.54   & 	0.017	\\	
        12					& 	  0.29   & 	0.01	\\\hline\hline	
	\end{tabular}
\end{table}

In addition to the outages caused by the concurrent LoRa (smart-meter) transmissions, we consider here outages from other (non-LoRa) users that use the same channel.
Different from \cite{Georgiou2017LowScale} and following \cite{Nardelli2016MaximizingConstraints}, we assume the noise is negligible compared to the interference.
If the non-LoRa devices' positions are modeled as a Poisson point process $\Phi_\mathrm{I}$ with density $\lambda_\mathrm{I}$ devices/km$^2$ and the aggregate interference is treated as noise by the gateway \cite{Nardelli2015ThroughputInformation}, the outage probability $P_\mathrm{out:2, x}$  in a link using a spreading factor $\mathrm{x}$ with $\mathrm{x} = 7,...,12$, with a respective SIR threshold $\beta_\mathrm{SFx}$, is given by: 
\vspace{2ex}
\begin{equation}
P_\mathrm{out:2, x} = 1 - \exp\left(-k \lambda_\mathrm{I} r^2 (\beta_\mathrm{SFx})^{2/\alpha}\right),
\label{eq:out2}
\vspace{1ex}
\end{equation}
where $\beta_\mathrm{SFx}$ is different for each SF (refer to Table \ref{tab:LoRa}).

\textbf{Remark:} Both \eqref{eq:out1} and \eqref{eq:out2} assume that all users transmit with the same power.
Different transmit power, or even some kind of channel inversion, may also be incorporated into the proposed formulation, as discussed in, for example, \cite{Nardelli2016MaximizingConstraints} and \cite{Haenggi2013StochasticNetworks}.
Although these differences would affect the overall link performance, it would not change its qualitative behavior in relation to the density of interferers and to the smart-meter-gateway distance, which are our focus in this paper.

The outage probability that the reference link experiences when their transmission occur using a spreading factor $\mathrm{x}$ can be computed as the complement of a successful transmission probability: $1- (1-P_\mathrm{out:1})(1-P_\mathrm{out:2, x})$.
Using  \eqref{eq:out1} and \eqref{eq:out2}, 
\begin{equation}
P_\mathrm{out, x}	= 1 - \exp\left(-k r^2 \left(\lambda  (\beta_\mathrm{SF})^{2/\alpha} + \lambda_\mathrm{I} (\beta_\mathrm{SFx})^{2/\alpha}\right)\right).
\label{eq:out-total}
\end{equation}

As the SF allocation is random and independent, the average outage probability $P_\mathrm{out}$ can be computed as follows:
\begin{equation}
P_\mathrm{out} = p_\mathrm{SF} \sum\limits_\mathrm{x}  P_\mathrm{out, x},
\label{eq:out-avg}
\end{equation}
remembering that, for LoRa, $\mathrm{x}=7,..,12$ and $p_\mathrm{SF} = 1/6$ when the allocation is uniform across the SFs.

Besides, each SF$\mathrm{x}$ has a different bit-rate $R_\mathrm{SFx}$, as shown in Table \ref{tab:LoRa},
This, however, does not include outages and it might be interesting to evaluate the effective bit-rate, here defined as $(1-P_\mathrm{out, x}) R_\mathrm{SFx, eff}$.
Similar to \eqref{eq:out-avg}, we can evaluate the average effective bit-rate $R_\mathrm{SF, eff}$ as:
\begin{equation}
R_\mathrm{SF, eff} = p_\mathrm{SF} \sum\limits_\mathrm{x} (1-P_\mathrm{out, x}) R_\mathrm{SFx, eff}.
\label{eq:rate-avg}
\end{equation}

\subsection{Numerical results}
\begin{figure}[!t]
\centering
	\includegraphics[width=0.9\columnwidth]{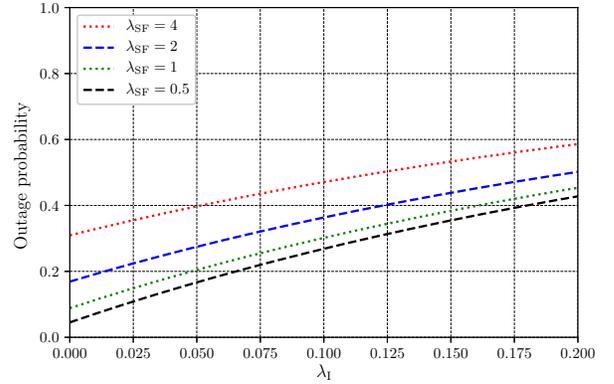}
	\caption{Average outage probability considering a random allocation of SF with probability $1/6$ as a function of the density $\lambda_\mathrm{I}$ of non-LoRa devices for different  densities  $\lambda_\mathrm{SF}$ of users served by the gateway and $r=1.5$km.}
	\label{fig:avg-out-li}
\end{figure}

\begin{figure}[!t]
\centering
	\includegraphics[width=0.9\columnwidth]{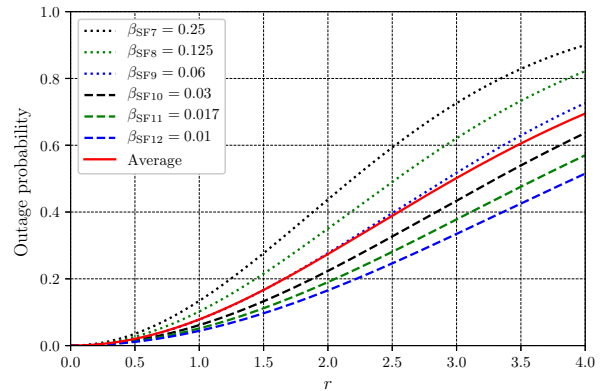}
	\caption{Outage probability vs. the distance $r$  for $\lambda_\mathrm{SF} = 0.5$ and $\lambda_\mathrm{I}=0.05$. The red curve is the average outage probability considering the smart-meters are  assigned to a specific SF with probability $1/6$.}
	\label{fig:out2-r}
\end{figure}

We present here the numerical results assuming: a path-loss exponent $\alpha = 4$ (urban environment), a SIR threshold related to LoRa end-devices using the same SF $\beta_\mathrm{SF} = 4$ (6dB) and a thinning probability $p_0 = 0.025$ related to the traffic and duty cycle constraint; the numerical setting related to each different SF is given in Table \ref{tab:LoRa}.
When not explicitly mentioned, we assume the distance between the reference smart-meter to the gateway to be $r=1.5$ km, while the density of active non-LoRa devices to be $\lambda_\mathrm{I}= 0.05$ devices/km$^2$.
{These two values were arbitrarily chosen since they do not qualitatively change the analysis \cite{Haenggi2013StochasticNetworks}; the way they affect the link performance will be presented when discussing Fig. \ref{fig:avg-out-li} and \ref{fig:out2-r}.}

\begin{figure}[!t]
\centering
    \includegraphics[width=0.9\columnwidth]{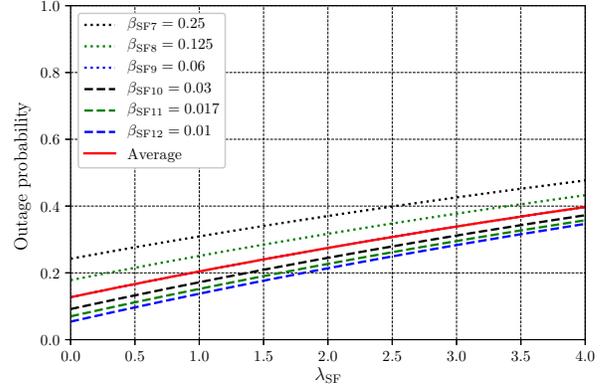}
	\caption{Outage probability as a function of the density $\lambda_\mathrm{SF}$ of users served by the gateway for the different spreading factors (SFs) and $r=1.5$km. The red curve is its average  considering $p_\mathrm{SF}=1/6$.}
	\label{fig:out-lsf}
\end{figure}

Fig. \ref{fig:out-lsf} shows how the density $\lambda_\mathrm{SF}$ of smart-meters affects the outage probability $ P_\mathrm{out, x}$ for each possible SF$\mathrm{x}$ and its respective average $P_\mathrm{out}$.
Although the relation between these variable shown by \eqref{eq:out-total} is exponential, the range of interest does not present a steep behavior, regardless of the SF.
This effect comes with the thinning processes related to the active concurrent transmissions at the same SF.
Therefore, the effects of the smart-meters' interference is not so dramatic.

\begin{figure}[!t]
\centering
	\includegraphics[width=0.9\columnwidth]{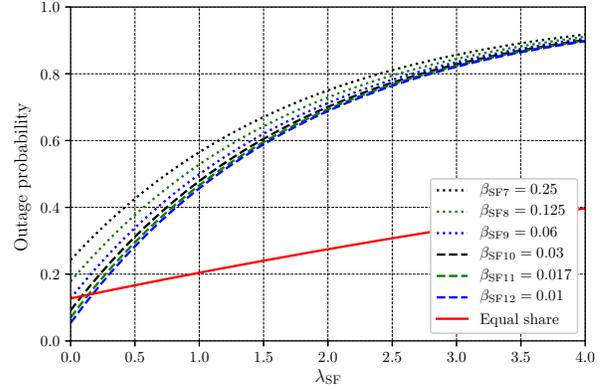}
	\caption{Outage probability as a function of the density $\lambda_\mathrm{SF}$ of smart-meters served by the gateway for the different SFs and $r=1.5$km. The red curve is its average considering $p_\mathrm{SF}=1/6$ while the other curves consider that all users are allocated only to one SF.}
	\label{fig:out2-lsf}
\end{figure}
\begin{figure}[!t]
\centering
	\includegraphics[width=0.87\columnwidth]{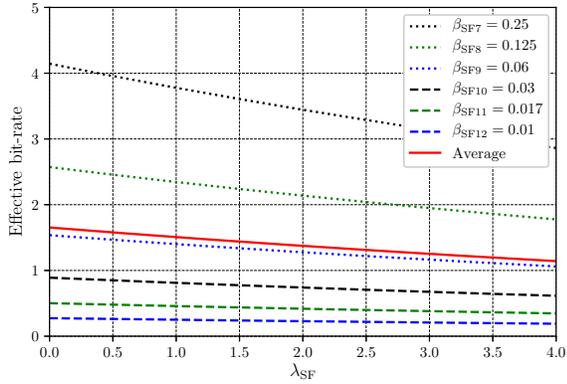}
	\caption{Effective bit-rate as a function of the density $\lambda_\mathrm{SF}$ of smart-meters served by the gateway for the different SFs and $r=1.5$km. The red curve is the average outage probability considering that the end-users are randomly assigned to a specific SF with probability $1/6$.}
	\label{fig:error-rate-lsf}
    \vspace{-1ex}
\end{figure}

Fig. \ref{fig:out2-lsf} reinforces this idea by showing the link outage probability when \textit{all} smart-meters are allocated in the same SF.
This case, compared with the average using an equal share (i.e. $p_\mathrm{SF}=1/6$, shows the steep exponential behavior so that the outage probability grows much faster with $\lambda_\mathrm{SF}$.
Besides, from both figures, we confirm that the lower the spreading factor, the higher the outage probability for the same setting.
This fact is expected since higher SFs imply  lower SIR thresholds for successful reception, obtained at expense of the link bit-rate.

Fig. \ref{fig:error-rate-lsf} shows the effective bit-rate for the different SFs and its average from \eqref{eq:rate-avg}.
As in Fig. \ref{fig:out-lsf}, one can see a smooth decrease (not steep as expected in exponential relations) in respect to $\lambda_\mathrm{SF}$, regardless of the SF.
The link can nevertheless transmit, in average, with a bit-rate between 1 and 2 kb/s.

\begin{figure}[!t]
\centering
	\includegraphics[width=0.9\columnwidth]{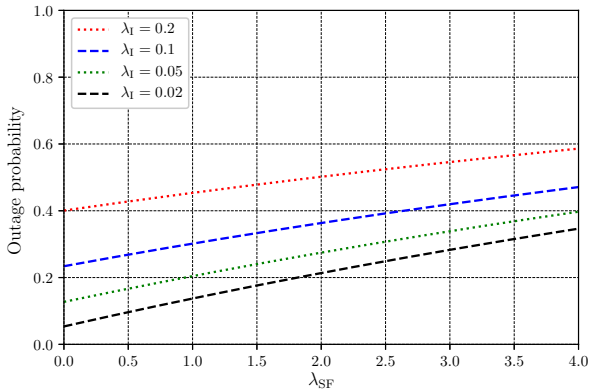}
	\caption{Average outage probability with a random allocation of SF with probability $1/6$ as a function of the density $\lambda_\mathrm{SF}$ smart-meters served by the gateway for different densities
    $\lambda_\mathrm{I}$ and $r=1.5$km.}
	\label{fig:avg-out-lsf}
\end{figure}

Figs. \ref{fig:avg-out-lsf} and \ref{fig:avg-out-li} present how the average outage probability changes with the density $\lambda_\mathrm{I}$ of non-LoRa devices.
For the range studied here, we can see that, in the worst case scenario with $\lambda_\mathrm{SF} = 4$ and  $\lambda_\mathrm{I}=0.2$ devices/km$^2$, the average outage is 60\% (a relatively high value).
Similar to the preset noise level in \cite{Georgiou2017LowScale}, the interference from non-LoRa devices imposes a floor level in the outage probability, but dependent on another parameter, namely $\lambda_\mathrm{I}$.
In specific terms, the higher $\lambda_\mathrm{I}$, the higher the outage floor, as indicated by \eqref{eq:out2}.
Consequently, if the density of non-LoRa devices in a given region is high enough, LoRa deployments will experience a very poor performance, specially for long range links.

To evaluate the effect of the distance on the link performance, Fig. \ref{fig:out2-r} shows the average outage probability when the distance between the reference smart-meter and the gateway changes.
As stated in \eqref{eq:out-total}, the relation is exponential and depends on $r^2$, so growing $r$ implies in a steep increase of outage events.
This plot is important for the network deployment when considering the sampling strategies since it helps to determine the worst case scenario so the most suitable position of the gateway can be chosen so as to achieve a minimum quality related to the signal reconstruction.

For example, if the gateway is planned to have a range of $r=4$ km, the worst case average outage probability is about $70\%$, which probably lead to a poor signal reconstruction (to be assessed in Section \ref{sec:sampling}).
This plot may also indicate that the SF allocation strategy used in \cite{Georgiou2017LowScale} based on distance ranges may outperform the random strategy.
However, from Fig. \ref{eq:out2}, this is not so obvious and requires further studies, since there is a clear trade-off involved between sharing the spectrum and the link distance.
This will be further discussed in Sec. \ref{sec:discussions}.

\vspace{-1ex}
\section{Sampling strategies}
\label{sec:sampling}
We followed here the results presented in \cite{Tome2016JointUsers}, but considering a different database \cite{Zimmermann2009End-useSavings}, which is comprised of around 400 houses with measurement lengths ranging from  a few days to a whole year with 10 minutes granularity.
The sampling strategies chosen are (i) time-based with a sampling frequency of 30 minutes, and (ii) event-based as described in \cite{Simonov2014HybridGrid}.
While the time-based implementation is straightforward, the other depends on some simple processing to identify a prescribed event.
Following \cite{Simonov2014HybridGrid,Tome2016JointUsers}, the event is defined by two situations: (a) a certain
amount of energy consumption $E_\mathrm{lim}$ is reached; or (b) a sudden change in the power demand denoted by $P_\mathrm{lim}$ is detected.

\begin{algorithm}[!t]
\caption{Event-based setting} 
\label{alg:event}
\begin{algorithmic}
\State $P_\mathrm{lim} \gets \mathrm{Set~Power~threshold}$
\State $P_\mathrm{step} \gets \mathrm{Set~Min~Power~Increase}$
\State $E_\mathrm{lim} \gets \mathrm{Set~Energy~threshold}$
\State $OK \gets $ False
\While{$OK$ is False}
\State $measEvent \gets EventMeasuring(P_\mathrm{lim}, E_\mathrm{lim})$
\If{$len(measEvent) \geq len(measTime)$}
	\State $P_\mathrm{lim} \gets P_\mathrm{lim}  + P_\mathrm{step}$ \Comment{Increase threshold}
	\If {$sum(measEvent('Power')) == 0$}:
    
    \Comment{Power threshold set too high}
    	\State $E_\mathrm{lim} \gets 2*E_\mathrm{lim}$ \Comment{Increase energy limit}
		\State $P_\mathrm{lim} \gets P_\mathrm{step}$ \Comment{Reset threshold}
	\EndIf
\Else
	\State $OK \gets$ True
\EndIf
\EndWhile
\end{algorithmic}
\end{algorithm}

\begin{figure}[!t]
\centering
    \includegraphics[width=0.95\columnwidth]{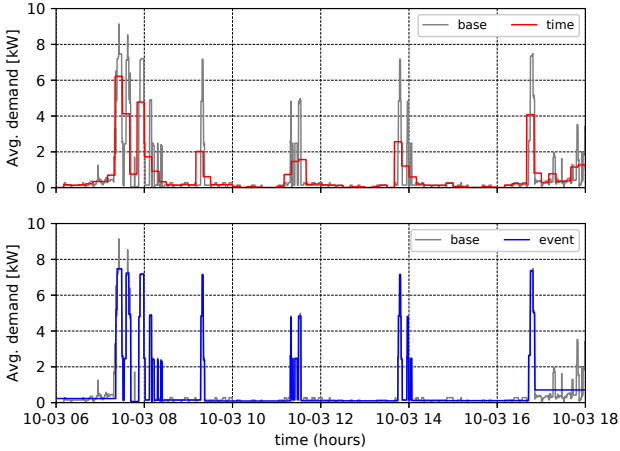}
	\caption{{Comparison between the actual measurements and sampled signals for a single house from 6:00 to 18:00 on October 3, 2016 ($x$-axis is presented as mm-dd hh). Top: Time-based; Bottom: Event-based.}}
	\label{fig:examples_sampling}
    \vspace{-3ex}
\end{figure}

Here, the initial parameters for the event-based approach were set as $E_\mathrm{lim} = 2$ kWh,  $P_\mathrm{lim} = 1$ kW, with increments of $P_\mathrm{step}= 0.5$ kW.
These values were arbitrarily chosen to lead to a smaller or equal number of samples compared to the time-based approach (48 samples per day). 
In the households that the event-based approach leads to more samples than the time-based, we implemented a simple procedure (presented in Alg. \ref{alg:event}) to modify the parameters in order to achieve a similar number of samples between the two cases.
Such procedure is neither optimal nor exhaustive; however, it builds a smaller set of measurements that is suitable for the present study.
On average, a reduction of about 17\% in the total number of measures was observed, but depending on the consumption patterns of the houses, the reduction ranges from about 70\% (14 measures per day on average) to 0 (no reduction). 

Fig. \ref{fig:examples_sampling} shows an example of the two strategies compared to the original measurements.
The results illustrate a single household between 6:00 and 18:00 on Oct. 3, 2006. 
Due to the thresholds chosen for plotting the event-based strategy, some smaller peaks were skipped (the power threshold is related to power variation, not to the absolute values). 
Remember also that the amount of points generated with these thresholds is designed to be similar to  the time-based strategy's amount.  

\begin{figure}[!t]
\centering
	\includegraphics[width=0.9\columnwidth]{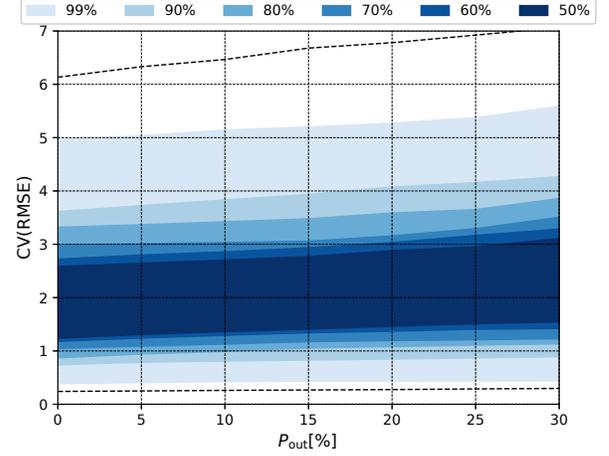}
	\caption{CV(RMSE) vs. the outage probability for the time-based strategy. Colored regions indicate the percentage of houses which belong to the range. Dashed lines indicate the extreme values.}
	\label{fig:error_time}
    \vspace{-1ex}
\end{figure}

\begin{figure}[!t]
\centering
	\includegraphics[width=0.9\columnwidth]{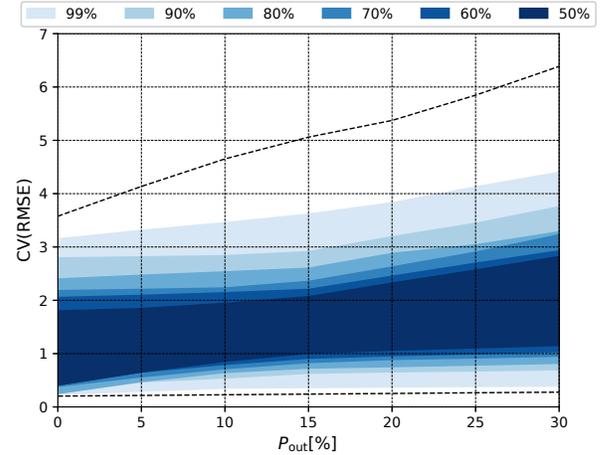}
	\caption{CV(RMSE) vs. the outage probability for the event-based strategy. Colored regions indicate the percentage of houses which belong to the range. Dashed lines indicate the extreme values.}
	\label{fig:error_event}
    \vspace{-3ex}
\end{figure}

We ran the same procedure illustrated in Fig. \ref{fig:examples_sampling} for 350 houses for a full week to assess how outage events from the LoRa system would affect the curve reconstruction  by the aggregator, following  \cite{Nardelli2016MaximizingConstraints,Tome2016JointUsers}.
The outage probabilities were varied from 0 to 30\% so the aggregator observes less points, which are lost with such a probability ({note that this value comes from the LoRa specification, as discussed in Sec. \ref{sec:system}).}
At the end of each day, the aggregator makes a linear interpolation between subsequent points to reconstruct the average power curve.
In the case that a sample is lost, the reconstruction error is computed as the relation between the interpolated point and the original (from the database).
The error analysis is based on 100 simulations for each household.
The quality metric was the root-mean-square error (RMSE) error between the measurements, normalized by the consumption of each one of the houses, herein called CV(RMSE).
While the measurements cannot be used to directly compare two houses with reasonably different consumptions, it provides a fair tool to compare the different sampling strategies {as well as compare the quality degradation due to different values of outage}.

Figs. \ref{fig:error_time} and \ref{fig:error_event} show the results for the time-based and event-based strategies. 
The color bands are the percentage of houses that fall into that range, whereas the dotted lines indicate the extreme values.
One can see that the time-based strategy is less sensitive to outage events, that is, its CV(RMSE) value has small changes as the outage probability grows, but its performance is overall poor when compared to the event-based strategy.
This indicates the time-based sampling is more redundant in many periods (e.g. during the night, when the electricity demand tends to be minimal) and missing points can be successfully reconstructed.
Conversely, the event-based samples tend to contain more information so missing points have a bigger impact in the signal reconstruction.

\begin{figure}[!t]
\centering
	\includegraphics[width=0.9\columnwidth]{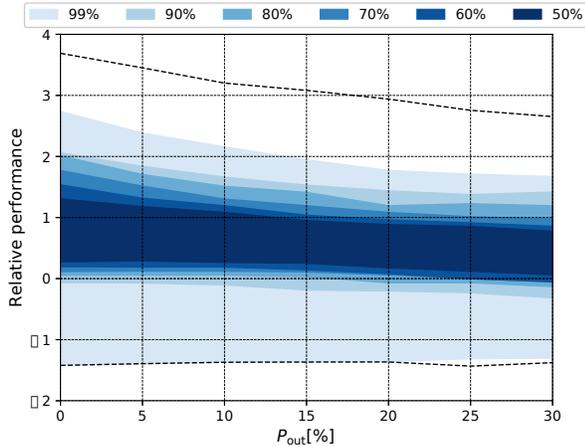}
	\caption{Relative performance between time-based and event-based strategies. Results above zero imply the event-based strategy is better (i.e. smaller error), while below zero the time-based strategy is better.}
	\label{fig:error_relative}
    \vspace{-2ex}
\end{figure}

Fig. \ref{fig:error_relative} compares the two strategies by subtracting the errors from both measurements to evaluate their relative performance. 
The positive values indicate that the event-based strategy outperform the time-based. 
It is possible to see that the event-based strategy provides better reconstruction in almost 90\% of the cases, regardless of the amount of error involved.
On the other hand, as previously discussed, this advantage tends to decrease if the outages become too frequent. 
It is important saying that this effect is a byproduct of the design of the event-based strategy where the number of samples cannot grow indefinitely, but is limited to about 48 per day (the number of time-based samples).
With such a limited number of samples, the information of each measurement grows.

\vspace{-1ex}
\section{Discussions}
\label{sec:discussions}
The results presented in the previous two sections showed that: (i) the outage probability is affected by the interference and the smart-meter-gateway distance and (ii) the event-based strategy usually outperforms the time-based in terms of reconstruction error. 
Now, let us consider the following situation: a communication engineer needs to deploy LoRa gateways in a city so that the reconstruction error is within a given quality limit.
From the proposed scenario, the engineer can only decide about the gateway range and the sampling strategy; the other parameters like density of smart-meters and non-LoRa devices, as well as the power demand, are given.

For example, a distribution company requires that the quality of the reconstructed signals by the aggregator has, in the worst case, a CV(RMSE) of 4 for the 90\% of the households under its coverage.
Using the event-based scheme, this performance can be achieved with an outage probability up to 30\%, as determined by Fig. \ref{fig:error_event}.
If the density of smart-meters is 0.5 and of non-LoRa devices is 0.05, then we can use the red curve from Fig. \ref{fig:out2-r} to find the gateway shall cover a range of $2$ km, at most.

This simple case illustrates how a LoRa communication system together with a ``smarter'' sampling strategy can be used to deploy electricity metering in cities.
Other important feature introduced by the event-based approach is an inherent peak detection in relation to variations in power demand, which may be useful to identify unusual or critical situations.
It is worth saying that, although the results are based on realistic numbers and on actual electricity demand data, they are used here to test a concept that must be further analyzed and optimized.
The same methodology may be directly applied to, for instance, water and gas metering since the consumption patterns are similar \cite{Mcneill2017EstimatingData}.
Another interesting use case might be related to commuting and traffic, whose daily patterns are somehow similar to electricity demand (e.g. \cite{Mcneill2017EstimatingData}); this, however, would require a more thorough study about the event definition and the gateway points.
In general, despite the particularities of each specific application, we understand that LoRa combined with an event-based sampling strategy provides a scalable solution for massive machine-type communications and IoT deployments needed in the future smart cities.

\section{Conclusions}
\label{sec:conclusions}
{This paper studied a LoRa wireless network deployment for electricity metering, where this technology combined with a event-based metering  strategy led to a fairly good quality signal reconstruction.
We presented here new results, which provide advances in the topic as follows.}

We extended the LoRa analysis of \cite{Georgiou2017LowScale} by using a stochastic geometry approach that includes LoRa and non-LoRa devices as interferers.
We employed a randomized (distance-independent) spreading factor allocation without favoring devices closer to the gateway, providing a fairer allocation for the specific metering application.
Note that other applications where fairness is not an issue would best suit with the usual distance-dependent allocation  (e.g. temperature monitoring in cities).
Therefore, we see our results as complementary to and consistent with \cite{Georgiou2017LowScale}.

{Our study also moved beyond \cite{Nardelli2016MaximizingConstraints,Tome2016JointUsers} by considering a LoRa technology in contrast to the previously developed cognitive radio approach based on abstract Shannon limits of interference-limited networks and perfect directional antennas. While \cite{Nardelli2016MaximizingConstraints,Tome2016JointUsers}, focused on optimizing the communication system performance under such idealized conditions, we assume LoRa specifications to serve as guidance for actual deployments. Nevertheless, although the present model and our objectives are quite different from [8], [9], the system performance (evaluated in terms of outage) is still limited by the same factor, namely co-channel interference.
}

{Besides, we reinforced the strength of the event-based metering introduced by Simonov et al. \cite{Simonov2014HybridGrid,Simonov2017GatheringMetering} (also used in \cite{Tome2016JointUsers}) by showing that a better signal reconstruction can be consistently achieved in comparison with the time-based one using a different dataset (i.e. different demand profiles and data granularity).
To reach this result, we proposed a simple algorithm that defines the events' thresholds from the consumption data (which are determined by each different household) so the number of samples generated by the event- and time-based are approximately the same. 
This algorithm is general and can be used to determine the thresholds based on historical data, and can be easily adapted to provide ``real-time'' adjustments.
}

{All in all, we argue that the proposed approach may be used in planning actual deployments by (\textit{i}) defining event-based metering using thresholds set based on historical data, and (\textit{ii}) defining LoRa gateways' locations since their range can be directly related to the signal reconstruction.
Although the results are quite abstract at this stage, we plan to develop this framework to real case-studies (ranging from dense cities to remote rural zones), as well as assess its feasibility to other smart city applications, as water metering and heating.
}

\bibliographystyle{IEEEtran}

\end{document}

%% file: lorafig.tikz
\centering
	\tikzstyle{quad}=[rectangle,
					thick,
					minimum size=2mm,
					draw=black,
					rounded corners=.1mm]
	\tikzstyle{quadmini}=[rectangle,
				minimum size=.4mm,
				draw=black,
				fill = red!40,
				rounded corners=0mm]					
					
	\tikzstyle{retang}=[rectangle,
					thick,
					minimum size=3cm,
					minimum height=1cm,
					draw=black,
					rounded corners=1mm]
	\tikzstyle{tri}=[isosceles triangle,
					isosceles triangle stretches,	                                    
					minimum width=1mm,
					minimum height=.5mm,
					inner sep=0pt,
					draw=black,
					shape border rotate=-90,
					thick]

\pgfmathsetseed{12345}
\begin{tikzpicture}[auto,node distance=3cm,>=latex]    


\draw[dotted, black] (0, 0) circle [radius=2cm]; 

\foreach \x in {1,2,...,4}{
	\draw[fill=blue!40] (.7+rand, .7+rand) circle [radius=.5mm]; 
	\draw[fill=blue!40] (-.7+rand, .7+rand) circle [radius=.5mm];
	\draw[fill=blue!40] (-.7+rand, -.7+rand) circle [radius=.5mm];
	\draw[fill=blue!40] (.7+rand, -.7+rand) circle [radius=.5mm];
}

\foreach \x in {1,2,...,4}{
	\draw[fill=black] (rand, rand) circle [radius=.5mm]; 
}

\foreach \x in {1,2,...,7}{
	\draw[fill=red!40,red!40] (.9+rand, .9+rand) circle [radius=.5mm]; 
	\draw[fill=red!40,red!40] (-.9+rand, .9+rand) circle [radius=.5mm];
	\draw[fill=red!40,red!40] (-.9+rand, -.9+rand) circle [radius=.5mm];
	\draw[fill=,red!40,red!40] (.9+rand, -.9+rand) circle [radius=.5mm];
}

\node [quad, name=source] (source) {}; 
\node [tri, name=s1, above of = source, xshift=0cm,yshift=-27mm] (s1) {};
\draw [draw=black,thick,-] (source) -- (s1);

\path[draw=gray,thick, dotted] (0,0) edge node [below]{$_{r}$} (2,0);

\draw[fill=black] (2, 0) circle [radius=.5mm]; 
\end{tikzpicture}